# Re-entrant Phase Transitions and Dynamics of a Nanoconfined Ionic Liquid

Stefano Mossa[*]

*Univ. Grenoble Alpes, CEA, CNRS, INAC-SyMMES, 38000 Grenoble, France*

Ionic liquids constrained at interfaces or restricted in subnanometric pores are increasingly employed in modern technologies, including energy applications. Understanding the details of their behavior in these conditions is therefore critical. By using molecular dynamics simulation, we clarify theoretically and numerically the effect of confinement at the nanoscale on the static and dynamic properties of an ionic liquid. In particular, we focus on the interplay among the size of the ions, the slit pore width, and the length scale associated to the long-range organization of polar and apolar domains present in the bulk material. By modulating both the temperature and the extent of the confinement, we demonstrate the existence of a complex reentrant phase behavior, including isotropic liquid and liquid-crystal-like phases with different symmetries. We show how these changes impact the relative organization of the ions, with substantial modifications of the Coulombic ordering, and their dynamical state. In this respect, we reveal a remarkable decoupling of the dynamics of the counterions, pointing to very different roles played by these in charge transport under confinement. We finally discuss our findings in connection with very recent experimental and theoretical work.

## I. INTRODUCTION

In ionic liquids (ILs), molecular structures that frustrate the growth of the crystal phase at room conditions interfere with electrostatic interactions between ions and generate a behavior which is often at variance with that of molecular liquids and liquid crystals [1,2]. On one side, this challenges in many ways our understanding of conventional electrolytes. On the other, it determines features which make ILs putative breakthrough materials for a few modern technologies, in particular for storage and conversion of energy [3–5]. In these applications, ILs are often in contact with interfaces [1,6] or confined in pores of subnanometric size [7], a situation that is highly nontrivial already for simple liquids [8]. A complete understanding of the physics of ILs in these configurations is therefore needed. Two examples, taken from recent experimental work on ILs confined at different length scales [9,10], illustrate the scope of the challenge.

Nanoscale capillary freezing at room temperature has been demonstrated for an IL confined between a fixed substrate and the tip of a tuning fork atomic force microscope, placed at a distance less than a well-defined threshold thickness [9]. This has been found to be as small as about 15 nm in the case of an insulating (mica) substrate and to significantly increase with the metallic character of the latter. Although challenged by the work of Ref. [11], these results point to a dependence of the phase behavior of constrained ILs on the nature of the confining interfaces, which determines the critical pore size.

Other remarkable properties have been discovered in even thinner metallic pores. An x-ray scattering and hybrid reverse Monte Carlo simulation study [10] has provided evidence of the existence of the so-called superionic state in ILs. This was predicted theoretically in Refs. [12–14], for pore widths comparable to the size of the adsorbed ions, in the subnanometric range. At these length scales, the ionic properties of matter are modified by the occurrence of an exponential screening of the ion-ion electrostatic interactions, due to the image forces induced at the pore walls. These enable ions of the same sort to preferentially fill the pore, substantially modifying Coulombic ordering and causing an anomalous increase of the capacitive properties [15,16], which is promising for supercapacitor technology [17–19]. Both of these examples ask for the comprehension of the role played by the size of the pore in determining the phase behavior of the adsorbed IL. Here, we address this issue systematically by molecular dynamics simulation, focusing on structure and dynamics of an IL confined in nanometric slit pores. These range from sizes of a few Å to 5 nm, encompassing the entire interval from extreme confinement, with pore widths reduced to a fraction of the ion size, to situations where a bulklike behavior is recovered. By also modifying the system temperature, we have sampled extensively the confined IL phase diagram. We present clear evidence that capillary crystallization is indeed possible for an IL confined at the nanoscale in an isolating pore.

[*]stefano.mossa@cea.fr

In addition, we demonstrate that the size of the (cat)ions, the width of the pore, and the length scale associated to the mesoscopic organization of ILs in the bulk [20] concur to determine a remarkable reentrant phase behavior. This includes different ionic-liquid-crystal phases together with liquid states, which develop into a gel phase at high temperature and extreme confinement. Based on an extremely extended set of data, we characterize the structural properties of the different phases, including the details of the mutual organization of the counterions, and elucidate the impact on the main features of the dynamics. Finally, we discuss our results with reference to recent experimental and theoretical work.

## II. RESULTS: MODEL AND LENGTH SCALES

Recent computational studies of ILs have often employed quite spherical representations for the cation, including the restricted primitive model of Ref. [13], the popular model for [$C_4$mim][$PF_6$] (1-butyl-3-methylimidazolium hexafluorophosphate) of Ref. [21], or even the all-atoms model for [$C_1$mim][Cl] of Ref. [22]. At variance with these works, we consider the IL [$C_{10}$mim$^+$][$PF_6^-$] (1-decyl-3-methylimidazolium hexafluorophosphate), with a strong molecular anisotropy. The anion $PF_6$ is represented as a single interaction center without internal structure, to exacerbate the differences with the cation. The latter, in contrast, is now a rodlike molecule of length $l_M \simeq 13.2$ Å, described by an optimized version of the coarse-grained model of Ref. [23] (see Fig. 7).

This choice makes it comfortable to highlight the interplay which involves $l_M$, the confining pore size $l_p$, and the length scale $l_Q$, associated to the mesoscopic nanostructuration of the bulk [20], which is particularly intense for ILs with long alkyl chains. [We can estimate $l_Q = 2\pi/Q^*$, where $Q^*$ is the position of the low-$Q$ prepeak appearing in the static structure factor $S(Q)$; see below.] Indeed, when $l_p \gg l_Q$, the pore width does not interfere with the density fluctuations associated to the above feature. As a consequence, we can expect the confined system to behave similarly to the bulk, and the properties of the adsorbed IL layer to be representative of those of the single interface case. In the opposite limit $l_p < l_M$, where the pore cannot accommodate a cation resting in the upward position, we should be able to probe extreme confinement conditions where the cations lay almost parallel to the interface with the pore, and the system can practically be considered two dimensional. In the intermediate range, for pore sizes of the order of a length scale $l_d$, associated to the wetting properties of the pore (see below), interesting features should finally appear as a consequence of the modifications induced by the pore on the IL density and orientational properties.

All the details of our reparametrization of the force field [23], molecular structures (Fig. 7), implementation of the unstructured slit pore (slab geometry in the $x$-$y$ plane), preparation of the adsorbed fluid, and clarifications on the numerical setup [we work in the $(NP_{x-y}T)$ ensemble] in comparison with the thermodynamic conditions met in experiments are deferred to Sec. VII.

## III. RESULTS: STATICS

### A. Thermodynamics

We have considered an extremely large data set of phase points (see Sec. VII), and we describe in detail the data at temperatures $T = 300$, 400, and 500 K. In Fig. 1(a), we plot the per-bead total potential energy $e_P = E_P/N$ of the IL beads as a function of $T$, at the indicated values of $l_p$. At $l_p = 50$ Å, we find a behavior $\propto T^{3/5}$, typical of molecular liquids in the bulk [24]. Surprisingly, at lower values of $l_p$ sudden drops appear at defined values of $T$, which depend on $l_p$ and point to the crossing of some form of phase boundary. At the lowest shown value $l_p = 10$ Å in the condition of extreme confinement, we recover a continuous dependence on $T$.

A similarly discontinuous scenario is found by quantifying the modifications of the IL mass density in our constant-$P_{x-y}$ conditions. In Fig. 1(b), we show $\rho(l_p)$ at the indicated values of $T$. At large $l_p$, $\rho$ decreases smoothly at all values of $T$. In the intermediate $l_p$ region, however, clear modulations develop, especially at the lower temperatures. In the inset, we show the data comprised in this region, now as a function of $l_p/l_M$. This representation is useful to highlight that, at $T = 400$ K, the extrema points of $\rho$ occur at $l_p/l_M \simeq 1$, 3/2, 2, therefore correlating to the formation in the pore of extremely ordered cation-rich layers, with a modulation which depends on the cation size. Note that, in conditions of extreme confinement, a deep drop in $\rho$ appears at around $l_p \simeq 7$ Å at all values of $T$, signaling the formation of a very low-density state. We will be more specific on this point below.

The observed modulations depend on the interaction of the IL with the pore walls. We have calculated $f_w$, the total interaction energy with the walls normalized to the pore surface [Fig. 1(c)]. At large $l_p$, $f_w$ assumes constant $T$-dependent values, until a $T$-independent crossover at $l_d \simeq 5/2 l_M \simeq 35$ Å. For lower values of $l_p$, the curves follow the increase of the confinement in a nontrivial fashion, with modulations similar to those of $\rho$ and wider at lower $T$. Note that $f_w$ is the analog of the force profiles measured by a surface force apparatus [25,26] and shows a behavior very similar to that of many confined ILs, pointing to strong layering effects in the pore. Details about the nature of this feature remain a lively subject of discussion [27].

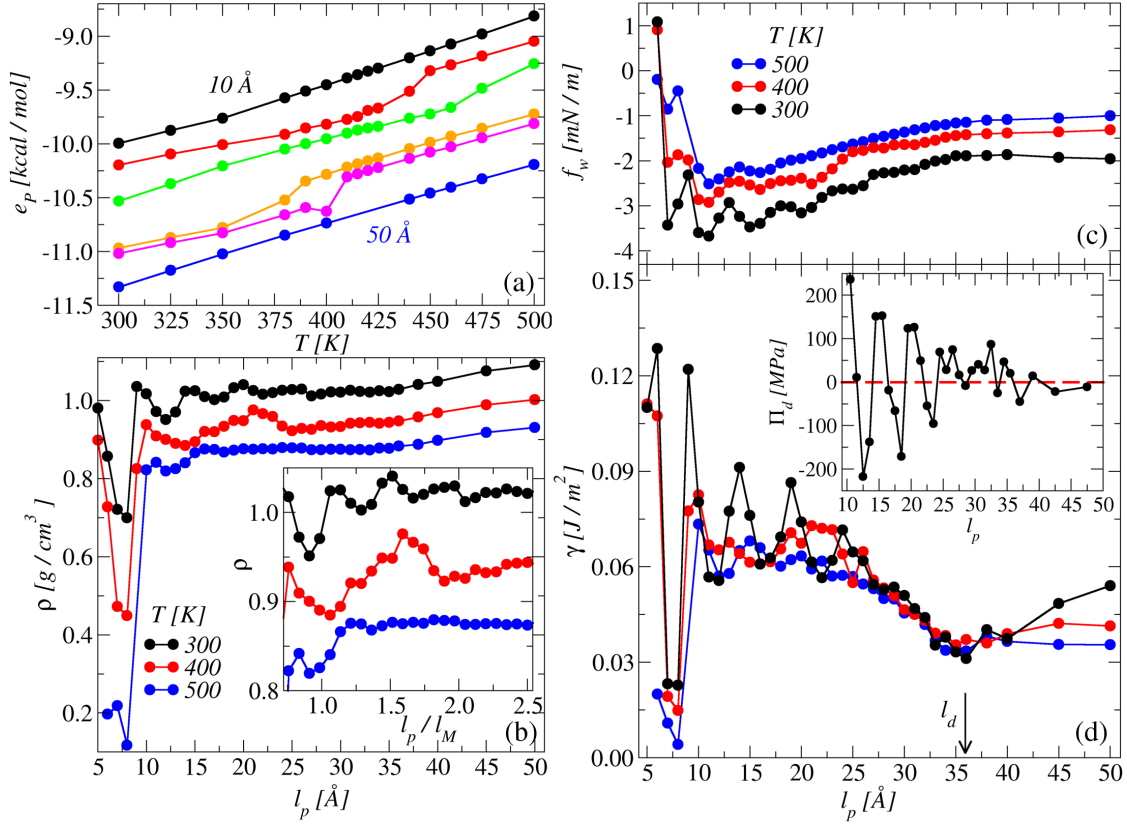

FIG. 1.  $T$ and $l_p$ dependence of thermodynamic observables. (a) Total per-bead potential energy $e_P$ as a function of $T$, for $l_p = 10, 12, 14, 20, 22, 50$ Å, from top to bottom. We observe a discontinuous behavior of the data at intermediate values of $l_p$, which points to the crossing of $T$- and $l_p$-dependent phase boundaries. (b) $l_p$ dependence of the IL mass density $\rho$ in the pore, at the indicated values of $T$. These data also show important discontinuities and point to substantial layering. In the inset we show the same data, now as a function of the pore width normalized to the length of the cation, $l_p/l_M$, to highlight the dependence on cation geometry at the different temperatures. (c) $l_p$ dependence of $f_w$, the energy of interaction with the pore walls normalized to the total surface of the interfaces, at the indicated values of $T$. (d) $l_p$ dependence of the surface tension $\gamma$ at the same values of $T$. These data allow us to be quantitative about the wetting properties of the absorbed liquid phases. We also show $l_d$, indicated by the arrow. In the inset, we show the disjoining pressure $\Pi_d$, calculated from the data for $T = 300$ K, as described in the text.

Important modifications also intervene in the wetting properties of the interfaces. We have calculated the surface tension as $\gamma(T, l_p) = l_p/2\langle P_{zz} - 1/2(P_{xx} + P_{yy})\rangle$, in terms of the diagonal components of the pressure tensor $P_{\alpha\beta}$. The data are shown in Fig. 1(d) as a function of $l_p$ at the indicated $T$, and they are intriguing. Indeed, at large $l_p$ and low $T$, they decrease smoothly, corresponding to the setup of a phase with wetting on the surface that is favored compared to the liquid, up to a $T$-independent pore size corresponding to $l_d$. Next, $\gamma$ undergoes an average $T$-independent increase by decreasing $l_p$ with, however, the presence of modulations whose width is especially important at low $T$ and which follow those of $f_w$.

Also, $\gamma$ is related to the total force exerted by the adsorbed fluid on the confining walls, the disjoining pressure [28], $\Pi_d(l_p) = -\partial_{l_p}\gamma(l_p)$, that we have calculated numerically and show in the inset for $T = 300$ K. $\Pi_d(l_p)$ features a damped oscillatory behavior [29], with a modulation dictated by the cation size. The oscillations are completely suppressed for $l_p \simeq l_d$, which can, therefore, be interpreted as the pore size at which the pressure normal to the wall tends to that in the bulk (at the same temperature and chemical potential) [29] and determines the infinite-pore limit. Overall these data unambiguously point to important phase modifications of the absorbed fluid following variations of both $l_p$ and $T$.

## B. Layering and orientational properties

The data presented above refer to observables averaged over the entire pore. They present, however, modifications that directly point to layering and, therefore, inhomogeneous space-dependent distributions of the different species in the pore that are worth studying. We have precisely identified the positions of the cation/anion-rich domains in the different conditions by additionally coarse-graining the cation in terms of the centers of mass of the polar (beads I1, I2, I3) and apolar (beads CM and CT) parts (see Fig. 7). We plot our $z$-dependent (one-dimensional) data for all investigated values of $l_p$ together in Figs. 2(a)–(c) (for the polar head and the apolar tail of the cation and the anion, respectively) at $T = 400$ K in the form of a color map. Note that we have rescaled the $x$-axis as $z/l_p$ in the range [−0.5, 0.5], and the $y$-axis as $l_p/l_M$, to underline the correlation of the appearing features with the steric constraints associated to the rodlike cations. Also, the contrast has been enhanced, rescaling the signal in the range [0, 1] in all cases (see color code).

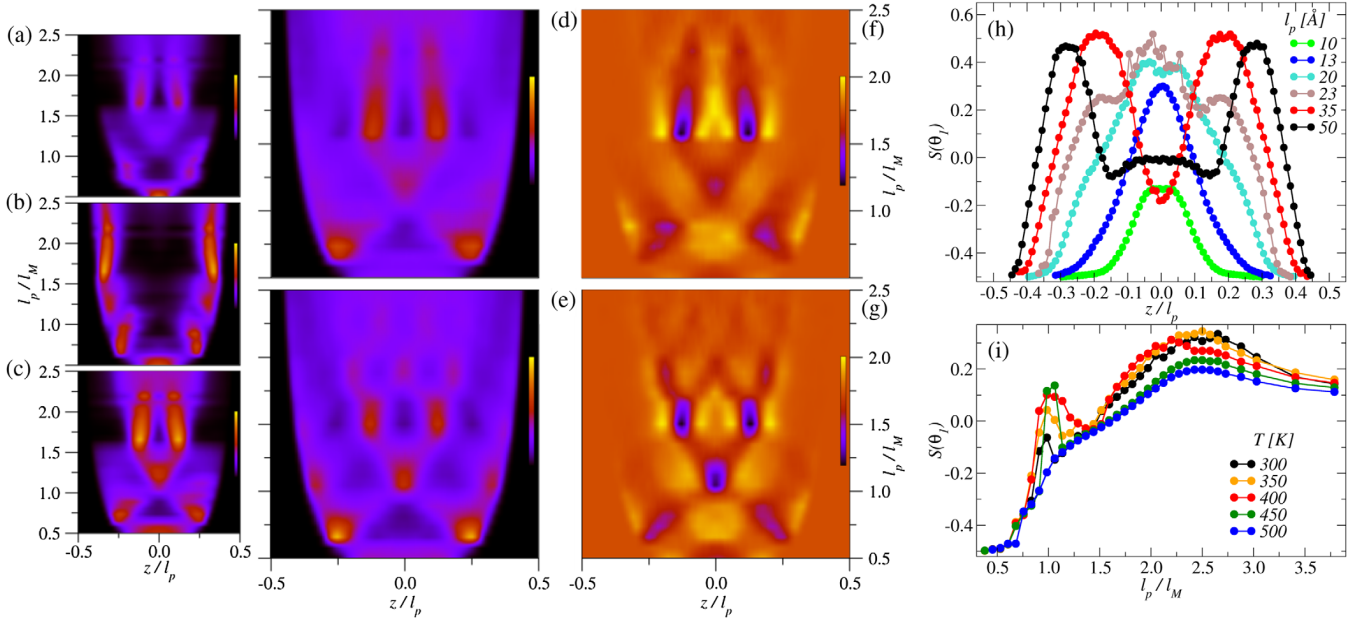

FIG. 2. Layering and orientational order. (a)–(c) Probability to find a bead associated to the polar and apolar parts of the cations and of the anions, respectively, at $T = 400$ K. The data are shown as a function of the position in the pore normalized to the pore size, $z/l_p$, for all values of $l_p/l_M$, and are renormalized in [0, 1] (see color box). (d),(e) Total mass distributions at $T = 400$ and 300 K, respectively, normalized in [0, 1]. (f),(g) Charge distributions at the same temperatures, normalized in the range [−0.5, 0.5]. (h) $z$ dependence of of the nematic order parameter $S(\theta_1)$ at $T = 400$ K at the indicated values of $l_p$. (i) Average values of $S(\theta_1)$ as a function of $l_p/l_M$ at the indicated values of $T$.

In conditions of extreme confinement, when the pore cannot accommodate a cation in the upright position, all species are unsurprisingly condensed in the center of the pore. In contrast, by increasing the pore width until $l_p \simeq l_M$, a clear partition occurs [30], with the apolar sections of the cations and the anions symmetrically localized at the interfaces, and the polar beads assuming a slightly more homogeneous structure shifted toward the center [31]. Interestingly, for $l_p > l_M$, the apolar tails coalesce at the interface, due to a value for $\epsilon_w$ [see Eq. (2)] which does not depend on bead type and to the absence of charged pore walls. (Note that this feature does not seem to play a significant role in strong confinement situations.) Concomitantly, the anion's probability first condenses in the center of the pore and next partitions into two well-defined regions symmetric to the pore axis. The polar heads of the cations follow a similar pattern.

This nonhomogeneous distribution of the different moieties in the pore induces strong space-dependent fluctuations in the mass distributions, which we plot in Figs. 2(d) and 2(e) at $T = 400$ and 300 K, respectively. (All data are normalized in the [0, 1] range.) Here, again, we find strong layering effects, particularly clear at $T = 300$ K, where a remarkable highly ordered three-layer structure is formed for $l_p/l_M \simeq 2$. This finding additionally underlines the importance of the steric constraints imposed by the elongated cations in fixing the periodicity of the modifications of the layered structure, while the actual distribution of mass and charge is primarily controlled by the anions. This last observation is corroborated by Figs. 2(f) and 2(g), where we now plot the charge distribution, normalized in [−0.5, 0.5]. We note immediately that the black (most negative) spots perfectly match the maxima in the mass distributions, with a complete ordering of the absorbed negative charge.

The above modifications are associated to space-dependent changes of the orientation of the cations. We have calculated the nematic order parameter, $S(\theta) = 1/2\langle 3\cos^2(\theta) - 1\rangle$, where $\theta$ is the angle formed by a director associated to the cation molecule with the normal to the pore plane. We have chosen as the directors both the normalized dipole moment with respect to the cation center of mass ($\theta_1$) and the normal to the plane identified by the three beads pertaining to the polar head ($\theta_2$). We show the $z$ dependence of $S(\theta_1)$ in Fig. 2(h), for $T = 400$ K at the indicated values of $l_p$. The data at $l_p = 50$ Å indicate important alignment (orthogonal) of the cations adsorbed at the interfaces, with the formation of an extended bulklike layer in the center of the pore, which disappears at 35 Å, giving way to non-negligible alignment in the pore plane. Interestingly, the formation of a three-maxima (positive) structure is detected on lowering $l_p$, until the formation of a unique layer in the center of the pore strongly aligned in the plane in extreme confinement conditions, as expected.

TABLE I.   Beads specifications including type, mass, charge, and associated neutron scattering length, in the indicated units.

| Bead type | $m_\alpha$ (g/mol) | $q_\alpha(e)$ | $b_\alpha$ |
|---|---|---|---|
| I1 | 35.5 | 0.356 | 6.23945 |
| I2 | 34.5 | 0.292 | 9.98035 |
| I3 | 26.0 | 0.152 | 5.815 |
| CM | 42.1 | 0.0 | −2.5002 |
| CT | 43.1 | 0.0 | −6.2411 |
| PF | 145.0 | −0.8 | 39.054 |

The $l_p$ dependence of $S(\theta_1)$ integrated over the entire pore is shown in Fig. 2(i), at the indicated values of $T$. The overall behavior of the data is qualitatively $T$ independent, except at the highest $T$. In extreme confinement conditions with $l_p \simeq 1/2l_M$, we find $S(\theta_1) = -1/2$, as expected for a dipole constrained to rotate in the plane of the pore. Consistently, $S(\theta_2) = 1$ (not shown), which corresponds to the imidazolium plane lying parallel to the pore walls. By increasing $l_p$, $S(\theta_1)$ first increases, reaching a positive maximum at exactly $l_p = l_M$ and pointing to some form of (smectic) order, as it will be clear below. Next, it decreases following an increase of orientational disorder associated to mass reorganization. The subsequent increase reaches a maximum for $l_p \simeq 5/2l_M$, corresponding to substantial nematic order, eventually tending to the isotropic value $S(\theta_1) \simeq 0$ [32].

## C. The static structure factors

The data discussed so far all point to a strong heterogeneity and nontrivial modifications of the structure of the absorbed IL, following a temperature-dependent interplay between the size of the rodlike cation and the width of the pore. We now focus on the correlations in space, which allow us to be more specific about the *nature* of the nanostructure of the system. We have calculated the (neutrons-weighted) static structure factors directly from the beads positions, as $S(Q) \propto 1/N\langle\sum_{ij} b_\alpha b_\beta \exp[i\mathbf{Q}\cdot(\mathbf{r}_i^\alpha - \mathbf{r}_j^\beta)]\rangle$. Here, $b_{\alpha(\beta)}$ is the coherent neutron scattering length for species $\alpha(\beta)$ (see Table I), calculated for united-atoms beads as the sum of the scattering lengths of its atomic components. $\langle\rangle$ indicates both the thermodynamic average and a spherical sampling over wave vectors of modulus $Q$ and polarized in the plane of the slab, $\mathbf{Q} = (Q_x, Q_y, 0)$.

Here, we discuss in detail the data at $T = 400$ K shown in Fig. 3(a), for the bulk and at the indicated values of $l_p$. In the bulk, we find the $S(Q)$ typical of ionic liquids with elongated cations [20], with the main diffraction peak at $\simeq 0.9$ Å$^{-1}$ and the remarkable prepeak located at $Q^* \simeq 0.25$ Å$^{-1}$. This feature corresponds to a typical length scale $l_Q = 2\pi/Q^* \simeq 25$ Å $\simeq 2l_M$, which can be associated to a mesoscale structuration of the three-dimensional network of the polar domains [33]. The prepeak is still present in mild confinement conditions corresponding to $l_p \simeq l_d$, which is too large to completely destroy the density fluctuations at $Q^*$, but is already able to strongly suppress its intensity.

Surprisingly, for $l_p = 23$ Å, the $S(Q)$ shows clear Bragg peaks superimposed onto a disorderedlike background, similar to that in the bulk, and also including a relic of the prepeak at $Q^*$. The Bragg peaks are produced by correlations of the charged species and are located at values

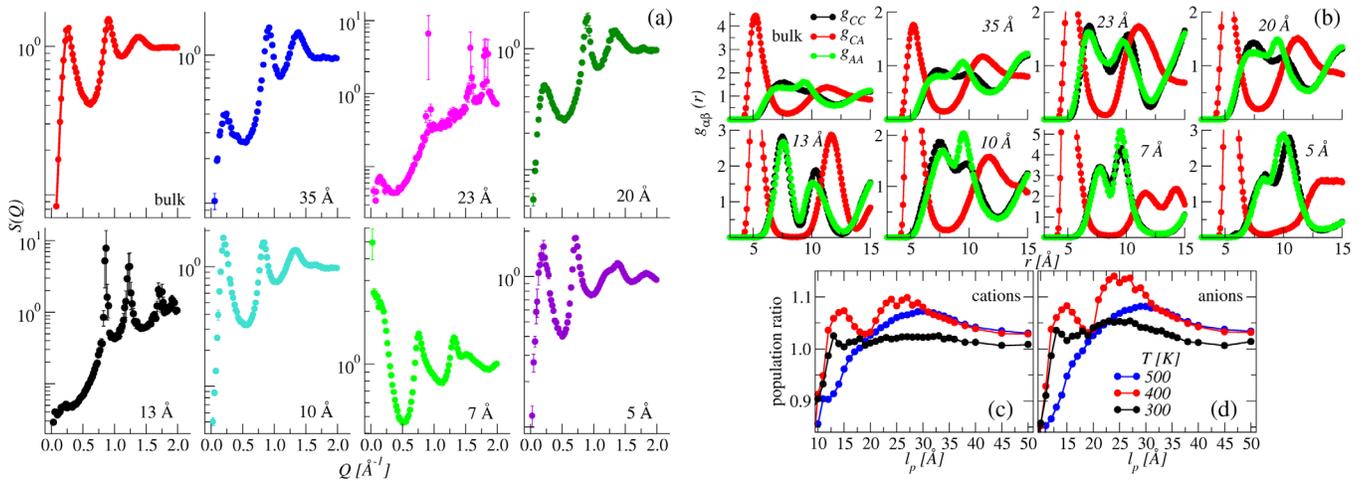

FIG. 3.   Details of the static structure and coordination properties. (a) Static structure factors $S(Q)$ in the bulk and at the indicated values of $l_p$, at $T = 400$ K. Data are shown on a logarithmic ($y$) scale. (b) Partial pair distribution functions $g_{CC}, g_{CA}, g_{AA}$ for the bulk and at the indicated values of $l_p$. (c),(d) $l_p$ dependence at the indicated temperatures of the fraction of coions in the first coordination shell normalized to the value in the bulk, for cations and anions, respectively.

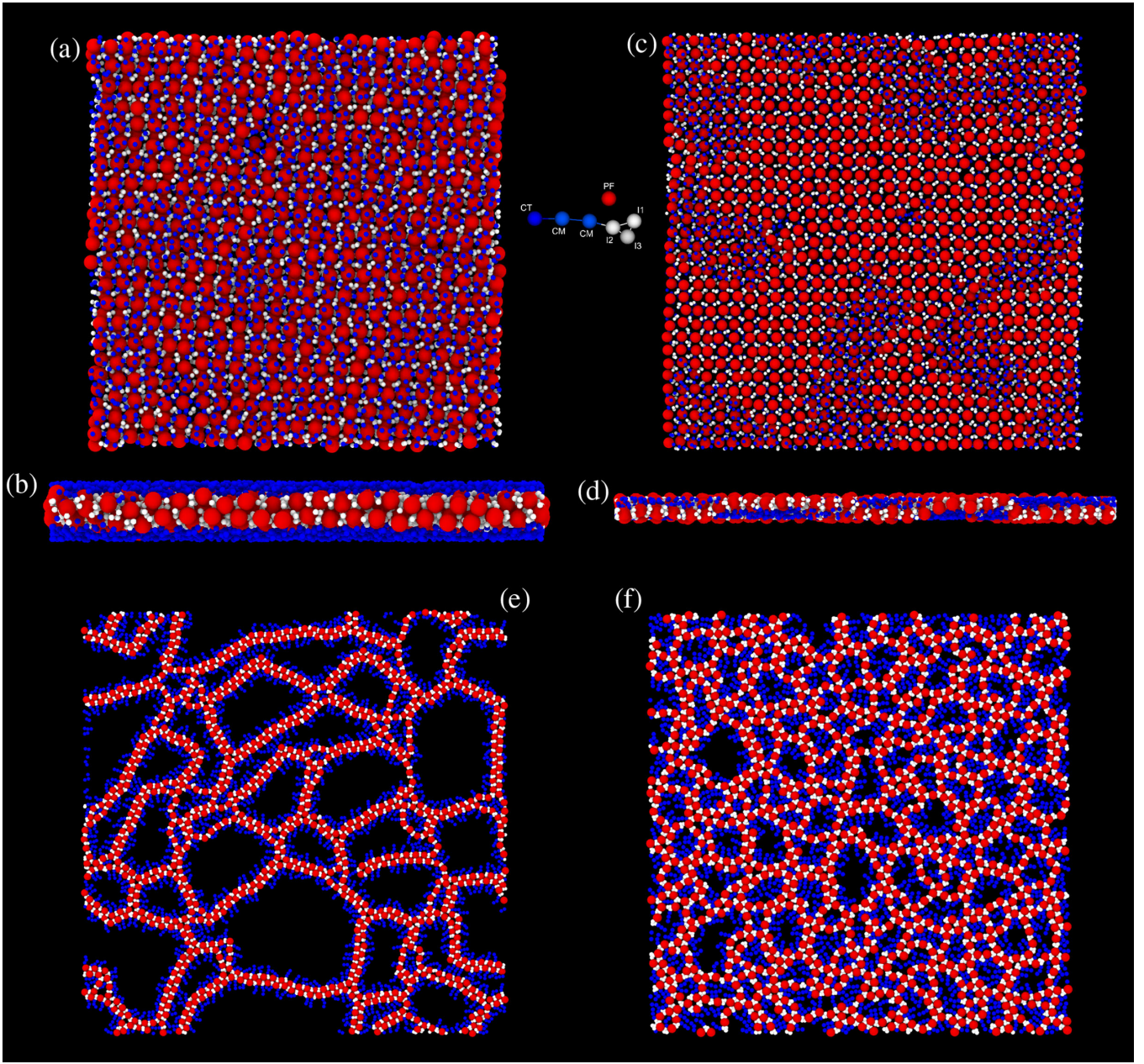

FIG. 4.   Typical systems snapshots at chosen $l_p$. We show typical configurations of the system in different conditions of confinement, at $T = 400$ K. Colors are the same as in the IL model shown in the same figure, and beads nomenclature can be found in the Methods section. Note that we have exaggerated the size of the (red) anions beads, to underline the symmetry of the ordered phases. (a),(b) Top and side view for $l_p = 23$ Å. (c),(d) Top and side view for $l_p = 13$ Å. In (e), we show a typical gel-like configuration at $l_p = 7$ Å modifying into a more homogeneous configuration in extreme confinement conditions at $l_p = 5$ Å, shown in (f).

of $Q_1 \simeq 0.9$ Å$^{-1}$ and $Q_2 \simeq 3^{1/2}Q_1 \simeq 1.5$ Å$^{-1}$, pointing to an ordered phase with a stringlike (or lamellar) symmetry. Strikingly, this structure is modified at lower $l_p$, where a pocket of liquidlike states is observed, before entering a new ordered region, with peaks now located at values of $Q_1 \simeq 0.85$ Å$^{-1}$ and $Q_2 \simeq 2^{1/2}Q_1 \simeq 1.2$ Å$^{-1}$, indicating a simple-cubic symmetry. When $l_p < l_M$, where the density of Fig. 1(b) shows a deep minimum, the $S(Q)$ diverges for

$Q \rightarrow 0$, pointing to gas-liquid coexistence. We will see in the following that an almost perfect network is formed in these conditions, similar to the case of an ionic liquid gel phase. At the lowest studied value of $l_p$, we recover again a homogeneous liquidlike state where, however, the prepeak is now almost as intense as the first diffraction peak, pointing to a substantially increased degree of order at long length scales, as expected in extreme confinement.

The above observations are confirmed by visual inspection of the systems snapshots in Fig. 4, where we also show the model for the IL as a reference for the color code used. Note that the size of the anions (red) beads has been substantially enhanced to emphasize the underlying order. The snapshots are shown both from the top of the confining surfaces and from a lateral view, for $l_p = 23$ Å [Figs. 4(a) and 4(b)] and 13 Å [Figs. 4(c) and 4(d)], respectively. The lamellar and cubic symmetries are very clear, with the lateral views emphasizing a substantial breaking of Coulombic ordering, as we will see in the next section. Ordering is even more clear on the bottom panels, where one can appreciate the formation of a gel-like network [Fig. 4(e)] and of a more homogeneous structure [Fig. 4(f)] in conditions of extreme confinement [34].

We have calculated the $S(Q)$ at all state points, which present features very similar to those discussed for $T = 400$ K, but corresponding to different values of $l_p$. In particular, data at $T = 300$ K are similar to those discussed above at large values of $l_p$, with modified locations of the transitions between phases at lower $l_p$, and without any intercalated liquid state. In contrast, at $T = 500$ K, the system keeps a liquidlike structure at all values of $l_p$.

We conclude this section with an observation which will prove useful in the final discussion. The systems with $l_p = 50$ Å should correspond to the case of a single interface, where a solid substrate is in contact with the *bulk* electrolyte. The properties of the interfacial layer in this configuration have been extensively scrutinized, both in experiments [36–38] and simulation works [39–41], and there is a quite general consensus that they can strongly differ from those of the bulk liquid. In particular, in the case of electrified interfaces, plain two-dimensional transitions to highly ordered phases can be induced at given applied potentials. In the case of dielectric substrates at zero potential, in contrast, the interface has been demonstrated to explore coexisting liquid and solid phases [37,40], showing an average net disordered character. Consistently with this last finding, from the analysis of the $S(Q)$, where the sum on beads was restricted to those lying in the first adsorbed layer extending for 7 Å from the pore walls (not shown), we have found a liquidlike scenario, without any sign of underlying order.

### D. The pair distribution functions

We now dig more into the relative organization of the ions. We have calculated the partial pair distribution functions $g_{\alpha\beta}$ with $\alpha, \beta = A, C$, considering the distances, $r_{ij}$, between the centers of mass of the polar head of the cations and the anions [42]. Our data are shown in Fig. 3(b), for $T = 400$ K at the indicated values of $l_p$. The structure of $g_{CA}$ shows no surprises, with an intense main peak centered at approximately 5 Å in all cases, which points to the formation of ion pairs, as is customary in ILs [13]. $g_{AA}$ and $g_{CC}$, in contrast, are more complex. In the bulk, $g_{CC}$ shows a broad peak intercalated to the two main peaks of the $g_{CA}$. $g_{AA}$ has a very similar shape, except for a tiny subpeak developing at less than 10 Å, which is more intense in mild confinement. At $l_p = 23$ Å, the situation changes drastically, with a clear splitting of the above feature into two well-defined subpeaks centered at approximately 7.5 and 10 Å for both cases, with a more intense peak at shorter distances. A similar behavior reappears in a few of the following panels, mirroring the apparently erratic behavior of the $S(Q)$ discussed above.

We can be more quantitative on these observations by calculating directly the coordination number, i.e., counting the number of coions and counterions comprised in the first coordination shell of any ion. This is defined as the sphere centered on an ion and of radius $r_c = 8$ Å, the position of the first deep minimum of the $g_{CA}$. From these data we can extract the fraction of coions over the total number of nearest neighbors, $p_{C,A} = n_{C,A}/(n_C + n_A)$, for the cations and anions, respectively, in analogy with Ref. [10]. In Figs. 3(c) and 3(d), we show $p_{C,A}$ normalized to the respective values in the bulk (the population ratio), as a function of $l_p$ at the indicated temperatures, which directly provides us with the relative variation of the Coulombic ordering compared to the unconstrained case.

At $T = 500$ K, $p$ slightly increases by lowering $l_p$, always staying above the bulk value and eventually decreasing for strong confinement. Surprisingly, at $T = 400$ K, the curve is strongly modulated, with clear maxima at defined values of $l_p$, and reaches values about 15% higher than in the bulk in the case of the anions. This effect vanishes at low temperature, where the value stays constant slightly above 1, until a very high extent of the confinement, where it eventually decreases. Typical systems configurations responsible for the maxima at $l_p = 23$ and 13 Å are those shown in the side views of Figs. 4(b) and 4(d). Note that these structures are very similar to the cartoons contained in Ref. [10], and our data point indeed to a facilitated approach in the first coordination shell of any ion of other ions of the same sort. This would imply important modifications of the Coulombic ordering present in the bulk, similar to the findings of Refs. [12,13]. We will address this point more in depth in the discussion section.

## IV. RESULTS: DYNAMICS

The analysis above points to strong discontinuous structural modifications of the system by changing both temperature and the size of the pore, compatible with plain phase changes in different confinement conditions. To complete the characterization of the system and provide an overall rationalization of the results, we now need to go deeper and clarify the modifications due to confinement on the dynamics of the ions. Here we focus on mass transport, with a Nernst-Einstein point of view. Future work will be devoted to clarifying the ionic properties of the system, which goes beyond the scope of this article.

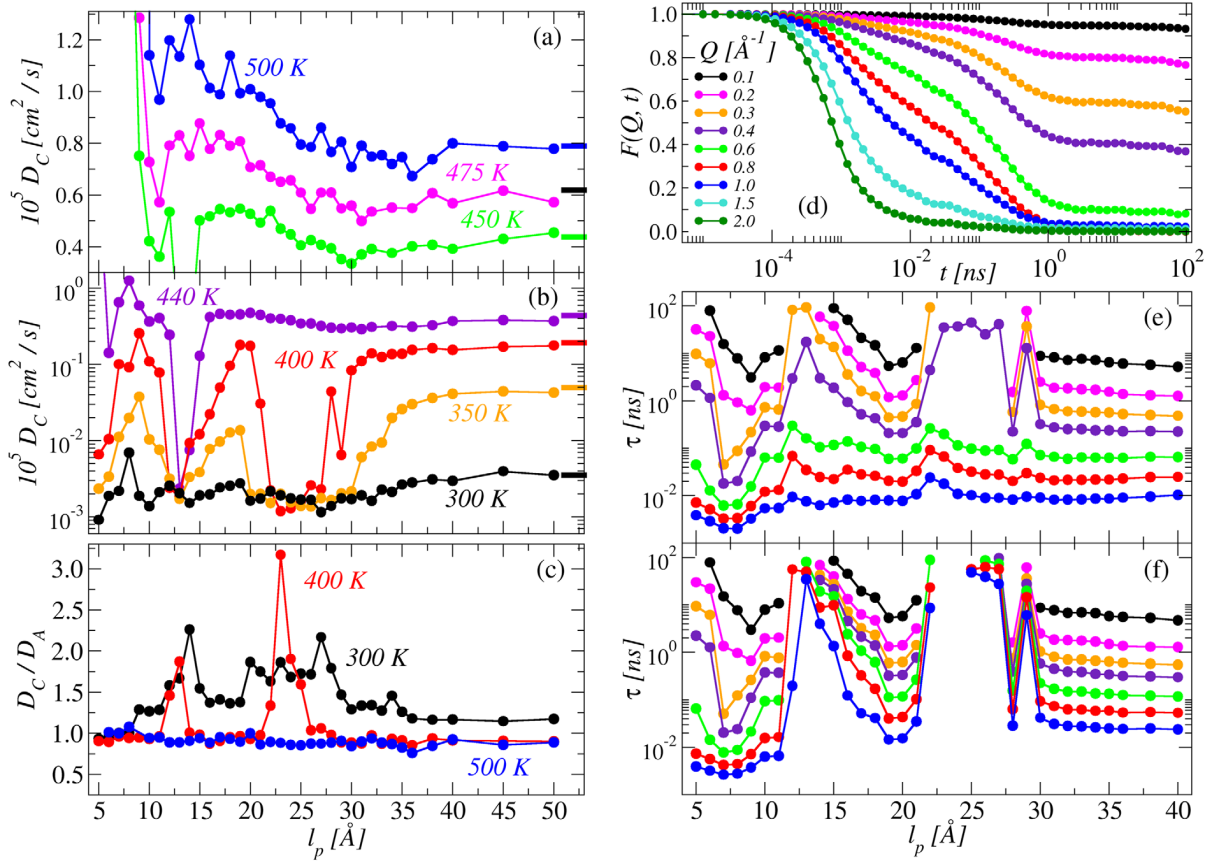

FIG. 5. Details of the dynamics. (a) $l_p$ dependence of the diffusion coefficients of the cations' centers of mass $D_C$ at the indicated high-temperature values. The corresponding values in the bulk are indicated by the solid thick lines on the right. (b) Same as in (a) at the indicated lower values of temperature. The deep discontinuities corresponding to phase-boundary crossing are discussed in the main text. (c) The ratios $D_C/D_A$ of the diffusion coefficients of the cations over those of the anions at the indicated temperatures. (d) Intermediate scattering functions $F_s(Q, t)$ at the indicated values of $Q$ and $T = 400$ K. The data are calculated for the cations' centers of mass. In (e) and (f), we plot the relaxation times $\tau(Q)$ for cations and anions, respectively, calculated from the relation $F_s(Q, \tau) = 1/e$. The colors identifying different values of $Q$ are the same as in panel (d). These data show that, for particular values of $l_p$, the dynamics of cations and anions decouples, the latter falling out of equilibrium on the timescale of our simulation, at all investigated values of $Q$.

## A. The diffusion coefficients

We start our analysis from the diffusion coefficients of cations and anions, extracted from the mean-squared displacements, $\langle R^2_{C,A}(t) \rangle = 1/N_{C,A} \sum_{i=1,N_{C,A}} |\mathbf{R}_i(t) - \mathbf{R}_i(0)|^2$. Here, $N_A = N_C$ is the number of cations and anions, and $\mathbf{R}_i(t)$ is the projection on the $x$-$y$-plane of the position vector of the center of mass of molecule $i$ at time $t$. Normally, the diffusion coefficient could be extracted from these data via the usual Einstein relation restricted to the $x$-$y$-plane. This relation, however, is only exact in the Fickian limit and is not appropriate in cases where, due to the confinement, strong subdiffusivity could appear. We have, in contrast, estimated $D = 1/4 \partial_t \langle R^2(t) \rangle$, from the local slope of the mean-squared displacement at long times. $D$ is, therefore, to be intended, more precisely, as an effective diffusion coefficient, characterizing the slow dynamics of the system on the timescale of our simulation, of about $10^{-1} \mu s$. Of course, at large $l_p$, we must recover a value close to that in the bulk. The data for the cations are reported in Figs. 5(a) and 5(b) at the indicated high and low temperatures, respectively. In all cases, we indicate the corresponding values in the bulk as solid thick lines, for reference.

Behavior at high and low temperatures differs drastically. At high temperatures $T > 440$ K, at large values of $l_p$, one recovers the value of the bulk, followed by a shallow minimum. At around $l_p \simeq l_d$, $D$ starts to increase and assumes values higher than in the bulk, at all $T$. This effect is not surprising as it has already been reported in experiments (see, for instance, Ref. [43] for [$C_8$mim][$BF_4$] confined in carbon nanotubes), and in the simulation work of Ref. [22] for [$C_1$mim][Cl], where it has been justified in terms of purely geometric arguments [44]. We are in the position to be even more specific on this point, by noticing that this increase mirrors perfectly an analogous increase in the surface tension [Fig. 1(d)], which can be rationalized as follows. In confinement conditions where the system stays in a stable liquid state, following an increase in surface tension, molecules will tend to stay away from the interfaces. They will, therefore, move toward the center of the pore, where the dynamics usually has a more pronounced bulklike character, with the net effect of increasing the average $D$ value.

At low temperatures $T \leq 440$ K, we observe a completely different behavior [Fig. 5(b)]. Indeed, the values of $D$ are always below the value assumed in the bulk, with the appearance of deep minima, at confinement sizes compatible with the discontinuities observed in the thermodynamical and structural properties. Note that the minima correspond to decreases of almost two orders of magnitude for both ions, providing an unambiguous evidence of the frozen character of the ordered phases described above. The data for the anions follow a qualitatively similar behavior, but an additional intriguing observation is in order.

We plot at each $l_p$ and at the indicated temperatures the ratio $D_C/D_A$ in Fig. 5(c). At the highest temperature, this ratio remains close to 1 in the investigated $l_p$ range [7]. By lowering $T$, however, corresponding to the discontinuities in $D_{A,C}$, the ratio increases up to a factor of 3 at $T = 400$ K. From these data, we conclude that, although, at the phase boundaries, both species are dramatically slowed down, their dynamics decouple. In particular, the anions properly freeze at the lattice site and, as a consequence, the cations settle in the remaining available volume. They are still able, however, to undergo nonvanishing local structural rearrangements (including intramolecular movements), probably sliding along the lattice planes identified by the anions ionic crystal [45]. Note that, especially in the case $T = 400$ K, the maxima correspond to those of Fig. 3(d), establishing a correlation between the dynamic decoupling and the partial breaking of the Coulombic ordering.

### B. The intermediate scattering functions

This scenario can be confirmed by looking at the $Q$-vector dependence of the self-intermediate scattering function, $F_s(Q, t) = 1/N_{C,A}\sum_{i=1,N_{C,A}} \exp\{i\mathbf{Q} \cdot [\mathbf{R}_i(t) - \mathbf{R}_i(0)]\}$. Here, we consider, again, the molecular center-of-mass position vectors and $Q$-vectors which are polarized in the plane of the pore. A subset of our data is shown in Fig. 5(d), for the cations at the indicated values of $Q$ and $T = 400$ K. The details of these curves are quite complex and a detailed analysis goes beyond the scope of the present work. We describe all data in a concise way in terms of a single structural relaxation time $\tau(Q)$, defined from the usual relation $F_s(Q, \tau) = 1/e$, avoiding any fitting procedure. The data are shown in Figs. 5(e) and 5(f), for the cations and anions, respectively.

For large values of $l_p$, the relaxation is constant at all values of $Q$, and very close to the bulk case for both ions. On lowering $l_p$, however, the relaxation time for cations exceeds the maximum timescale probed in our simulations ($\tau > 0.1\ \mu s$) at small values of $Q$ entering our accessible time window for larger values of $Q$, well above the value corresponding to the long-range structuration in the bulk.

This is in contrast with the case of the anions, where the relaxation exceeds our time window at all values of $Q$, pointing to completely frozen density fluctuations. A similar scenario appears again at lower values of $l_p$, where, however, relaxation timescales are reduced at high-$Q$ values. Note that the two frozen regions are separated by deep minima, corresponding to the intercalated liquid pockets, as expected. Eventually, in a situation of extreme confinement, we find again a substantial decrease of the relaxation time associated to the formation of the gel and the almost two-dimensional liquid described above.

## V. RESULTS: THE $(T - l_p)$ PHASE DIAGRAM

Based on the thermodynamic data, static structure factors, direct visual inspection of system snapshots, and dynamical behavior, we distill all the information in the $(T - l_p)$ phase diagram of Fig. 6. We also show the data in terms of $l_p/l_M$ (right axis), to help highlight the interplay with the cation molecular length. In the high-$l_p$ region for $l_p \geq l_d$, the confined ionic liquid is very similar to the bulk liquid at all temperatures (L), both regarding structure and dynamics, with the typical low-$Q$ prepeak in the $S(Q)$. Indeed, the extent of the confinement is still not sufficient to interfere destructively with the density fluctuations corresponding to this feature, the only visible effect being a decrease of its intensity.

The structure does not change much even at low $l_p$ and $T = 500$ K, where, however, a certain degree of cation nematic order sets in for $l_p < l_d$. Here, we have demonstrated that the interaction force between the confining walls starts to be non-negligible, with the typical oscillatory behavior of the disjoining pressure setting in. This point also corresponds to an increase of the diffusion coefficients of both ions compared to the values in the bulk, which we have explained in terms of reinforced interaction of the adsorbed fluid with the pore walls. The situation does not change for lower temperatures, $T > 420$ K, where, however, the system starts to cross the boundary of the phase $S_1$ at around $l_p/l_M \simeq 1$, eventually phase separating in extreme confinement, by entering the ionic liquid gel region ($G$).

For $T < 420$ K and $l_p < l_d$, the system starts to probe a remarkable reentrant phase behavior, freezing in the two solid phases $S_2$ (lamellar order) and $S_1$ (cubic order). We have shown that the symmetry of the two solids is determined by the freezing of the (spherical) anions at the lattice sites, with the cations occupying the interstitials with substantial smectic order induced by the pore interfaces. Also, note that the two solid domains are centered around $l_p/l_M = 1$ and 2, additionally highlighting the role played by the geometry of the cation in determining the position of the different boundaries in the phase diagram. We add, for completeness, that the boundaries separating $S_1$ and $S_2$ from the liquid apparently both have a discontinuous

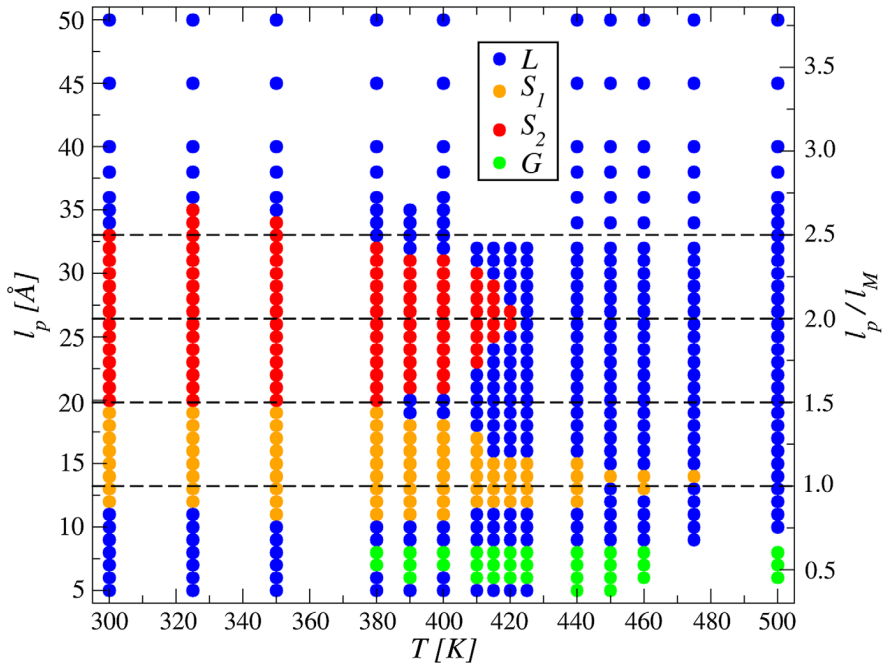

FIG. 6. Phase behavior of the confined IL. The $(T - l_p)$ phase diagram, on the basis of the discussed structural and dynamical properties. We also show the data in terms of $l_p/l_M$ (right axis). This is the most important result of the present work, demonstrating the possibility of capillary freezing of an ionic liquid in a nanopore, the existence of multiple arrested states, and the occurrence of a reentrant phase behavior with the stabilization of liquidlike pockets in situations of severe confinement. An interesting gel-like phase is also found, in conditions of very low density and high temperature, for confinement widths less than the rodlike cation molecular length.

character, unlike the finding of Ref. [47], for the case of confined water films.

The two solids are separated by an intercalated region of liquid pockets, which stay stable for $T > 380$ K. At $T = 380$ K and $l_p \simeq 20$ Å, we, therefore, find a solid-solid-liquid triple point, from which a solid-solid transition line originates that develops into the entire explored low-temperature region. Finally, for $l_p < 3/4 l_M$, a liquid region opens, which is stable at low temperatures but is destabilized in favor of an ionic gel phase at higher $T$.

We conclude the section with two remarks. First, a detailed analysis of the nature of the observed transitions or a precise determination of the phases' coexistence regions would involve much more refined numerical techniques than those employed here, including free energy calculations [48]. Note, for instance, that the reported phase diagram does not show even schematically the (very thin) liquid-solid coexistence lines. Being aware of the difficulties to be precise on this point, we have more easily opted to directly assign points showing clear signs of phase coexistence to the ordered phase, on the basis of the first occurrence of well-defined Bragg peaks. Second, the presence of electric fields in the pores, as natural in applications, significantly modifies the composition of the adsorbed layers and the interaction of the ions with the surface. The computed phase diagram may, therefore, drastically change under such conditions. Both these observations trigger interesting perspectives for this work.

## VI. DISCUSSION

We discuss our results in relation with previous experimental and theoretical work. Some aspects of the structure/dynamics interplay we have described in this work can be *generically* identified in other systems. For instance, the reentrant behavior in the phase diagram of Fig. 6 is similar to that demonstrated in Ref. [49], using density-functional theory for a model liquid crystal in a slit pore. A reentrant melting transition preventing solid-to-solid transformations and giving rise to intervening liquid phases has also been reported in Ref. [50] for the case of hard spheres in soft confinement. Finally, a reentrant behavior of solid phases with intercalated liquid pockets has been discussed in the case of confined water nanofilms in Ref. [47]. In contrast with that paper, however, we find no sign of a continuouslike phase transition, with abrupt changes of the diffusion coefficients when crossing both boundaries following temperature changes. Features of the dynamics reminiscent of those described here have also been reported. As an example, on one side the effect of layering on the diffusion coefficient in the plane of the pore, with the alternation of maxima and minima, has been highlighted for the confined liquid films of Ref. [51]. On the other side, the nonmonotonic dependence of the diffusion coefficient on the pore size has been reported for the confined hard spheres systems of Ref. [52], in the context of the glass transition phenomenon.

More *specifically*, it is instructive to discuss our results compared to two recent important pieces of work. First, nanoscale capillary freezing of ionic liquids confined between metallic interfaces has been demonstrated in Ref. [9], based on atomic force microscope (AFM) measurements. Although the investigated IL ([C$_4$mim][BF$_4$]) was different than that we have employed here, we can speculate that the experiment probed the equivalent of the capillary freezing at $T = 300$ K from the liquid to the ordered phase described above. Differently than here, however, freezing was detected for confinement lengths substantially larger than our $l_d \simeq 3.5$ nm ($\lambda_s = 15 \pm 6$ nm for mica). This discrepancy can be due to the accuracy of our molecular modeling approach; differences in the experimental setup compared to the simulation (as discussed in Sec. VII); or the effect of the dynamics of the AFM tip, a feature absent in our treatment, where the pore boundaries are both immobile. Also, note that, in the same paper, the crucial role played by the metallic character of the confining substrate was demonstrated, another ingredient which is lacking in our calculations. These are, however, in agreement with the computer simulation used to back up the experimental data (see the Supporting Materials of Ref. [9]), which demonstrated a shift in the melting temperature controlled by the pore diameter by using insulating confining boundaries. We can conclude that the metallic character of the pore certainly strongly enhances the occurrence of capillary freezing. This, however, seems to be also possible in insulating pores, at the expenses of even more extreme confinement conditions.

We close the discussion on this point by noticing that the conclusions of Ref. [9] have been challenged by measurements performed by means of a dynamic surface force apparatus [11]. In this work, the argument of the formation of solid prewetting films on the substrate was raised as an alternative to the genuine crystallization triggered by confinement at the nanoscale of Ref. [9]. In Sec. III C, however, we have specified that we did not detect any form of order in the first layer adsorbed at a single interface. This clearly points to the absence of any prewetting, while capillary crystallization is certain. Our work, hence, seems to strengthen the interpretation of Ref. [11], although it is not possible at this stage to be final on this issue.

Second, confined ILs are of increasing interest in energy applications, including supercapacitors technology, due to peculiar capacity effects (see, among others, Ref. [53] or [17–19]). These appear in conditions of strong confinement similar to that considered here, and they are rationalized in terms of the insurgence of a superionic state [13]. The possible significance of our work for this issue is worth discussing. The formation of a superionic state has been predicted as a direct consequence of screening the electrostatic interactions between ion pairs in extreme confinement in metallic pores. Very recently, the existence of the superionic state has also been validated experimentally [10] in terms of significant breaking of the Coulombic ordering, based on diffraction data and reverse Monte Carlo methods. The similarity of the data of those papers with our results included in Figs. 3(c) and 3(d), or with the system configurations of Figs. 4(b) and 4(d), looks surprising if one recalls again that we have considered insulating boundaries. Here, the partial breaking of Coulombic ordering appears as the natural consequence of mass and charge distribution in the pore, controlled by the steric constraints associated to the cation molecular structure, and without any relation to charges induced at the boundaries.

We add that our simulations, for the first time to the best of our knowledge, also provide a complete coherent picture of both phase behavior and consequent dynamics changes of the adsorbed IL. In particular, we have shown that corresponding to the modifications of the charge distribution under confinement, one can observe a remarkable decoupling of the dynamics of cations and anions. In our model, these two features are, therefore, correlated, another aspect shared with the superionic state [6]. Deeper insight, however, is needed to be conclusive on this point and attack other issues, as the modifications of ion conductivity beyond the Nernst-Einstein picture. We note, in conclusion, that our work highlights a possibly very different behavior of an IL adsorbed in nanometric pores of the same size, but at different temperatures. This poses a serious caveat on the legitimacy of extrapolating, at room conditions, properties observed at higher temperatures, a procedure that is very popular in simulation works.

## VII. METHODS

### A. The coarse-grained model for the IL

We have slightly reparameterized the model of Ref. [23], which reproduces nicely the static properties of room-temperature ionic liquids of the family 1-n-alkyl-3-methylimidazolium hexafluorophosphate, but also shows a partially satisfying dynamics, as noticed in Ref. [54]. More precisely, we have aimed at reproducing, in the investigated $T$-range, realistic time scales for the diffusion of both ions, mimicking at the same time an interesting feature observed in numerous ILs with cations of length $\lesssim l_M$ (see, for instance, Ref. [55]). In these systems, in fact, the cation diffuses faster than the anion, a feature not expected on the basis of simple ionic size considerations and which can be attributed to the more dynamically heterogeneous local environment of the former.

The cation atoms are coarse grained in beads, closely following Ref. [23]. As shown in Fig. 7, the imidazolium ring is composed of three charged beads, dubbed $I1$, $I2$, and $I3$. The uncharged alkyl chain, whose length can be tuned to reproduce the entire [C$_n$mim] family, is formed in the present case [C$_{10}$mim] by two CM beads terminated by a CT unit. We have estimated the molecular length $l_M \simeq 13.2$ Å, by measuring the distance of the two farthest beads $d_{I1-CT}$ in the completely relaxed configuration of the isolated molecule. The anion is, in contrast, a simple pointlike interaction center, PF. The values for the beads' masses and charges, together with the neutron scattering lengths used for the calculation of the static structure factor, are reported in Table I.

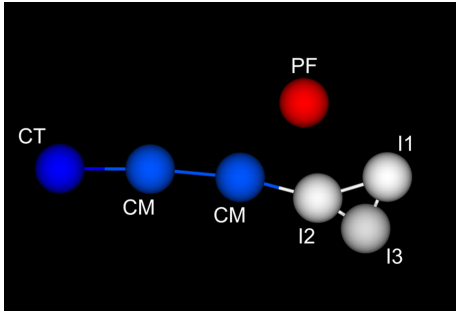

FIG. 7. Schematic of the coarse-grained model for $[C_{10}mim]^+[PF_6]^-$ of Ref. [23], where the reader can find the mapping of atoms to the coarse-grained beads. The cation polar head is a flexible structure formed by the (white) charged beads I1, I2, I3. The apolar tail is a linear chain constituted by two CM beads, terminated by CT (all in blue). The anion is a pointlike interaction center, PF (red). Parameters for the cation intramolecular potential are those of Ref. [23], and the optimized values for the pair interactions are reported in Table II.

TABLE II. Force-field parameters used in the simulations. We have considered the values in Ref. [23], by doubling the value of $\epsilon^{PF-PF}$, with $\sigma^{PF-PF}$ unchanged. Next, we have globally rescaled all $\sigma$'s by $f_\sigma = 0.948$ and all $\epsilon$'s by $f_\epsilon = 0.95$ and recalculated the interaction's energy with PF by the usual mixing rule $\epsilon^{\alpha\beta} = (\epsilon^{\alpha\alpha}\epsilon^{\beta\beta})^{1/2}$.

| Beads pair | $\epsilon$ (kcal/mol) | $\sigma$ (Å) |
|---|---|---|
| I1-I1 | 0.3569 | 3.8868 |
| I1-I2 | 0.3424 | 3.8394 |
| I1-I3 | 0.2603 | 3.8394 |
| I1-CM | 0.3309 | 4.0764 |
| I1-CT | 0.3718 | 4.1143 |
| I1-PF | 0.6178 | 4.8348 |
| I2-I2 | 0.3284 | 3.8868 |
| I2-I3 | 0.2497 | 3.8394 |
| I2-CM | 0.3174 | 4.0764 |
| I2-CT | 0.3566 | 4.1143 |
| I2-PF | 0.5926 | 4.8348 |
| I3-I3 | 0.1898 | 3.4128 |
| I3-CM | 0.2413 | 3.5645 |
| I3-CT | 0.2711 | 3.6024 |
| I3-PF | 0.4505 | 4.5978 |
| CM-CM | 0.3990 | 4.2717 |
| CM-CT | 0.4218 | 4.3091 |
| CM-PF | 0.6532 | 4.6480 |
| CT-CT | 0.4455 | 4.3466 |
| CT-PF | 0.6903 | 4.6859 |
| PF-PF | 1.0694 | 4.9296 |

The interaction potential is nonpolarizable, a limitation common in these studies (see Ref. [56] for a rapid review of modeling of ionic liquids). The nonbonded interaction between two beads $i$ and $j$ of type $\alpha$ and $\beta$, placed at a distance $r_{ij}^{\alpha\beta} = r^{\alpha\beta}$, is the sum of a 9–6 Lennard Jones potential and a Coulomb term,

$$V(r^{\alpha\beta}) = \frac{27}{4} \epsilon^{\alpha\beta} \left[ \left( \frac{\sigma^{\alpha\beta}}{r^{\alpha\beta}} \right)^9 - \left( \frac{\sigma^{\alpha\beta}}{r^{\alpha\beta}} \right)^6 \right] + \frac{1}{4\pi\epsilon_0} \frac{q^\alpha q^\beta}{r^{\alpha\beta}}. \quad (1)$$

The new values for the parameters are reported in Table II. All values of $\sigma^{\alpha\beta}$ have been rescaled by the same factor $f_\sigma = 0.948$ compared to those of Ref. [23]. The $\epsilon^{\alpha\beta}$ have been corrected in the same fashion, by a factor $f_\epsilon = 0.95$. In contrast, in order to cure the undesired features of the dynamics, we have substantially increased the strength of the PF − PF interactions, which now is $\epsilon^{PF-PF} \simeq 1.07$ kcal/mol. All other interaction energies involving the PF bead have been generated by the usual mixing rule, $\epsilon^{\alpha\beta} = (\epsilon^{\alpha\alpha}\epsilon^{\beta\beta})^{1/2}$. Following these modifications, we did not need to tune the values of formal charges (see, for instance, Ref. [21]), which are, therefore, the same as in Ref. [23]. In Eq. (1), the interactions are cut off (but not shifted) at $r = r_C = 15$ Å, while the wave-vector part of the long-range Coulomb interactions in the slit pore (slab geometry) are calculated via the adapted method of Ref. [57], as we will see below. The values of the parameters for the harmonic bonds, $V_B(r) = k_B(r - r_o)^2$, and angles, $V_A(\theta) = k_A(\theta - \theta_o)^2$, are the same as in Ref. [23].

## B. The confining pore

We consider a planar capillary (slit pore) with periodic boundary conditions in the $x$-$y$ plane. Confinement in a slab oriented with the normal in the $\hat{z}$ direction and width $l_p$ is implemented by bounding the simulation domain with two continuous unstructured flat walls located at $z = \pm l_p/2$. Each wall interacts with a bead at $z_i = z$ by generating a force in the $\hat{z}$ direction, corresponding to a potential [22],

$$V_w(z) = \epsilon_w \left[ \frac{2}{15} \left( \frac{\sigma_w}{z} \right)^9 - \left( \frac{\sigma_w}{z} \right)^3 \right] \theta(z - z_w). \quad (2)$$

We have chosen $\epsilon_w = 0.2$ kcal/mol, $\sigma_w = 3$ Å, and $z_w = 2.5\sigma_w = 7.5$ Å, for all bead types. Therefore, for $l_p < 2z_w = 15$ Å, the beads interact with both walls. Also, note that, here, the focus is on the direct effect on the system behavior of the length scale of confinement, rather than that of the nature of the interaction with the pore boundaries. The actual value of $\epsilon_w$ has, therefore, been simply fixed on the basis of previous work, and with the main requirement to be of a strength comparable to the pairwise interaction energy scales (excluding the anion-anion ones). We have checked that larger values, up to approximately $\epsilon_w = 1.6$ kcal/mol, do not seem to qualitatively change the phase behavior of the system. Obviously, we are aware of the fact that avoiding considering a more realistic confinement, for instance, by a crystalline lattice formed by immobile charged atoms, imposes limitations on our analysis (we cannot study capacitive features, for instance).

Care must be taken for the calculation of the interaction with the charge images in the nonperiodic $\hat{z}$ direction. This is done by treating the system as if it were periodic in $\hat{z}$, but inserting empty volume between atom slabs and removing dipole interslab interactions so that slab-slab interactions are effectively turned off. The methodology behind the algorithm used is explained in Ref. [57]. Note that this procedure corresponds to an insulating pore configuration.

### C. Simulation details

All calculations have been performed with LAMMPS [58]. In all cases, we have considered $10^3$ ion pairs, for a total number of interacting beads $N = 7000$. The trajectories have been propagated in the $(NP_{x\text{-}y}T)$ ensemble, where $L_z = l_p$ was fixed, while $L_x = L_y$ were allowed to fluctuate to match the value $P_{x\text{-}y} = 1$ atm. We have generated extremely extended data sets encompassing 15 values of temperatures in the range 300–500 K and confinement lengths in the range 5–50 Å. This amounts to an over-sampled state space including a total of 488 points, which are completely independent by construction. The simple fact that this very large ensemble of uncorrelated systems finally organizes on the completely consistent and neat phase diagram of Fig. 6 points alone to the strong statistical significance of the data.

Initial systems configurations were prepared as follows. We started from a completely randomized configuration at $T = 500$ K in a simulation box extremely elongated in the nonperiodic walls-bounded $\hat{z}$ direction, determining approximate Coulomb interactions truncated and screened according to the modified Wolf version of the damped shifted force model [59], with a damping parameter $\alpha = 0$. Next, we generated all pores with different widths by rapidly rescaling $L_z$ to the desired values of $l_p$. We have checked that all initial configurations did not present any form of order.

Each system was subsequently allowed to evolve, thermalizing on timescales of the order of $10^2$ ns at the lowest temperatures, followed by production runs of $10^2$ ns for all states. The total simulation time exceeds the $2 \times 10^2$ ns in most cases. With this procedure and having avoided, in particular, to prepare initial configurations by compressing/dilating or cooling systems generated in different conditions, we are confident we have minimized possible correlations among pores. In a few relevant cases, we also evolved multiple systems corresponding to the same $l_p$, but starting from different completely randomized initial configurations and following different independent thermal paths. We subsequently verified that the final systems had equilibrated in the same ordered structure. Note that the reported ordered configurations formed completely, in all cases, within an initialization run of 5 ns. This amounts to structures which stay stable on timescales even 40 times longer than the initial self-organization time, pointing to a quite robust stability of the reported phase behavior.

In addition, at each investigated value of $T$, we have studied the corresponding system in the bulk, with the same number of ion pairs, same values of the parameters involved in the simulation algorithms, and on the same timescales. We can, therefore, count in all cases on the exact reference bulk configurations. More details on the bulk systems will be discussed in forthcoming publications.

We conclude with an observation on our simulation's setup compared to the typical thermodynamic conditions met in experiments. Here, we have worked in thermodynamic conditions $P_{xx} = P_{yy} = 1$ atm, which are not those found in experiments, where, in contrast, the chemical potential $\mu$ is held constant. The typical setup corresponds indeed to the triple point conditions, where the pore is in contact with a liquid phase region surrounded by its vapor, and these two buffer phases keep $\mu$ constant in the pore. This layout can certainly be implemented in simulations (as, for instance, in Ref. [29]), but it is computationally more demanding than that we have employed here. At this point, every state we have investigated is, in principle, at a different value of $\mu$, making it arduous to propose a simple mapping to realistic thermodynamic conditions for an experiment aiming to demonstrate the reentrant behavior, for instance. We note, however, that, in Ref. [60], it was shown that solvation forces, which play a crucial role in these systems, are not very sensitive to changes in the chemical potential of the adsorbed fluid. This is reassuring in view of the relevance of our simulation for realistic experimental conditions.

### ACKNOWLEDGMENTS

We are indebted to Benoit Coasne for a few enlightening discussions, and thank Patrick Judeinstein for a careful reading of the manuscript. We acknowledge technical support from Olivier Blondel and Christine Leroy of the Information Technology Service (STI) at CEA Grenoble, and financial aid from the CEA funding program "DSM Energie".